# Spin-transport, spin-torque and memory in antiferromagnetic devices: Part of a collection of reviews on antiferromagnetic spintronics

J. Železný,[1][*] P. Wadley,[2][*] K. Olejník,[3] A. Hoffmann,[4] and H. Ohno[5, 6, 7, 8]

[1)]Max Planck Institute for Chemical Physics of Solids, Nöthnitzer Str. 40, 01187 Dresden, Germany

[2)]School of Physics and Astronomy, University of Nottingham, University Park, Nottingham NG7 2RD, United Kingdom

[3)]Institute of Physics, Academy of Sciences of the Czech Republic, Cukrovarnicka 10, 162 00 Praha 6, Czech Republic

[4)]Materials Science Division, Argonne National Laboratory, Argonne, IL 60439, USA

[5)]Center for Spintronics Integrated Systems, Tohoku University, 2-1-1 Katahira, Aoba-ku, Sendai 980-8577, Japan

[6)]Center for Innovative Integrated Electronic Systems, Tohoku University, 468-1 Aramaki Aza Aoba, Aoba-ku, Sendai 980-0845, Japan

[7)]Laboratory for Nanoelectronics and Spintronics, Research Institute of Electrical Communication, Tohoku University, 2-1-1 Katahira, Aoba-ku, Sendai 980-8577, Japan

[8)]WPI Advanced Institute for Materials Research, Tohoku University, 2-1-1 Katahira, Aoba-ku, Sendai 980-8577, Japan

[*]These authors contributed equally

**Ferromagnets are key materials for sensing and memory applications. In contrast, antiferromagnets that represent the more common form of magnetically ordered materials, have so far found less practical application beyond their use for establishing reference magnetic orientations via exchange bias. This might change in the future due to the recent progress in materials research and discoveries of antiferromagnetic spintronic phenomena suitable for device applications. Experimental demonstrations of the electrical switching and electrical detection of the Néel order open a route towards memory devices based on antiferromagnets.  Apart from the radiation and magnetic-field**

**hardness, memory cells fabricated in antiferromagnets are inherently multilevel which could be used for neuromorphic computing. Switching speeds attainable in antiferromagnets far exceed those of the ferromagnetic and semiconductor memory technologies. Here we review the recent progress in electronic spin-transport and spin-torque phenomena in antiferromagnets that are dominantly of the relativistic quantum mechanics origin. We discuss their utility in pure antiferromagnetic or hybrid ferromagnetic/antiferromagnetic memory devices**

# Introduction

Magnetism and technology have been inextricably linked for centuries - from the navigation compass, to motors and generators, to magnetic data storage. The field of information storage has been dominated by ferromagnetic materials from its inception to the present day. Memories based on ferromagnets, such as magnetic tapes and hard disk drives, have been one of the key factors that enabled the information revolution with its ubiquitous access to data everywhere at any time. Antiferromagnetism, on the other hand, has played fleeting roles in the story of information technology so far, despite the fact that antiferromagnetic order is common in magnetic materials. Most of these roles have been as passive elements, such as for pinning or hardening of ferromagnet layers.

The major obstacle which has kept antiferromagnets away from applications is that they are hard to control and hard to read. While ferromagnetic order can be detected by the magnetic fields it creates, and in turn manipulated by an external magnetic field, antiferromagnets produce no fringing magnetic fields and are much less sensitive to them (although they can also be manipulated by large enough magnetic fields). So while antiferromagnets could be used for memories, just like ferromagnets, the difficulty of detecting and manipulating the antiferromagnetic order provided a seemingly insurmountable barrier.

While magnetic fields provide a practical way for detecting and manipulating ferromagnetic order, many other methods have been developed. Perhaps most importantly, electrical currents can now be used both for detection and switching of ferromagnetic order. Utilizing electrical currents instead of magnetic fields is more efficient and more scalable, and thus the latest magnetic random access memories rely entirely on electrical currents. [1] The possibility of using electrical currents instead of magnetic fields for detection and manipulation has inspired a renewed interest in antiferromagnetic materials. Electrical manipulation combined with electrical detection of antiferromagnetic order has been recently demonstrated. [2] This shows that antiferromagnets could be used to store information in electronic memory devices and also opens new avenues in the fundamental research of antiferromagnetic order and dynamics.

In this article we review recent theoretical and experimental progress on spin-transport and spin-torque phenomena allowing for reading and writing information stored in antiferromagnets. See Fig. 1 for a summary of all proposed electrical reading and writing methods. We also discuss other transport phenomena relevant to spintronics such as the generation of spin currents by antiferromagnets due to the spin Hall effect (SHE). [3,4]

## Nonrelativistic spintronics effects

The giant [5,6] and tunneling magnetoresistance [7,8] and the spin-transfer torque [9, 10, 11] are among the key spintronics phenomena for applications in ferromagnetic memories. These effects occur in devices composed of two ferromagnetic layers separated by a thin metallic or insulating layer. The structures are made such that the magnetization of one ferromagnetic layer is free to rotate while the other is fixed (usually by coupling to an antiferromagnet). The metallic devices are called spin valves, while the tunneling ones are termed magnetic tunneling junctions. The giant and tunneling magnetoresistance are respectively the dependence of the ohmic or tunnel current on the relative orientation of magnetizations of the two ferromagnetic layers. This enables detection of the magnetization reversal of the free layer. These magnetoresistance effects can be very large; the tunneling magnetoresistance can be hundreds of percent even at room temperature.

When an electrical current is flowing through the structure it transfers spin from one layer to the other, which generates the spin-transfer torque. For a sufficiently strong current, this torque enables switching of the free magnetic layer between parallel and antiparallel configurations. The spin-transfer torque together with the tunneling magnetoresistance are used to create the basic building blocks of magnetic random access memories. [1]

Initial theoretical research in antiferromagnetic spintronics focused on antiferromagnetic analogues of spin valves and tunneling junctions (see Fig. 1a for illustration of two different states of such junctions). Núñez et al. [12] predicted that in an antiferromagnetic spin valve, a giant magnetoresistance as well as a current-induced torque will occur. Note that the current-induced torque is also sometimes referred to as the spin-transfer torque. It is not necessarily related to global spin transfer as highlighted in the calculations for antiferromagnets, [12] but can still be described in terms of local spin transfer. The mechanism in antiferromagnets differs from ferromagnets since the electrical current in (collinear) antiferromagnets is not spin-polarized. Núñez et al. [12] illustrated the existence of magnetoresistance and current-induced torque on a 1D model. Following works found these effects also in realistic antiferromagnetic spin valve structures [13,14, 15] and in tunneling junctions . [16,17] (See also reviews by MacDonald et al. [18] and Haney et al. [19] on the early calculations and comparison to ferromagnets.) Haney et al. [20] have shown that an antiferromagnet can also generate a torque on a ferromagnet. Similarly to ferromagnets, a current-induced torque is also predicted to occur for an antiferromagnetic domain wall [14, 21, 22]. All of these works, however, considered ballistic transport in perfectly epitaxial and commensurate heterostructures. A key issue is whether these effects will survive in realistic devices. In most of the calculations, it was found that phase coherence plays a crucial role and thus the presence of inelastic scattering is expected to be detrimental. This was indeed confirmed by Duine et al. [23], who have calculated the effect of inelastic scattering on the giant magnetoresistance and the torque. They found that these effects are much more sensitive to the presence of inelastic disorder in antiferromagnetic spin valves than in ferromagnetic ones. Nevertheless, it was argued that the magnetoresistance and the torque should survive since the inelastic mean free path is typically relatively large and elastic scattering was assumed to not influence the magnetoresistance and the torque much. Consequent works [24,15], however, studied theoretically the effect of elastic scattering on the torque in a metallic spin valve and found that the torque is very strongly suppressed even by the elastic scattering (the giant magnetoresistance was not studied in these works however). A. Manchon found that in antiferromagnetic spin valves no torque is present in the semiclassical limit [25]. Thus the torque in antiferromagnetic spin valves is probably limited to very clean samples and low temperatures. Recent calculations by Saidaoui et al. [26] showed that the elastic scattering has a much smaller effect on the torque in antiferromagnetic tunneling junctions than in the metallic spin valves. Therefore using tunneling structures might be experimentally more feasible than using the metallic spin valves; however, more research is needed to understand precisely the influence of disorder.

The role of the epitaxy of the antiferromagnetic heterostructures on the magnetoresistance and the torque has not been studied in depth. However, it is likely that a very good epitaxy is necessary. As can be seen in Fig. 1a, the two states of a simple antiferromagnetic junction differ only by a shift of one of the antiferromagnetic layers. Thus the presence of steps on the interface would destroy both the magnetoresistance and the torque (this is also reflected in the strong dependence of the torque on spacer thickness [24]). Therefore, while theoretical works have clearly demonstrated that giant and tunneling magnetoresistances and current-induced torques can in principle exist in purely antiferromagnetic structures, the observation of these effects would require very clean and perfectly epitaxial devices. Monolayer step edges are extremely difficult to avoid and make such devices very challenging to fabricate. This is in contrast to ferromagnetic spin valves and tunneling junctions, which are much more robust. The sensitivity of the magnetoresistance in antiferromagnetic spin valves is in agreement with the fact that experimentally only very small magnetoresistance in metallic [27,28] and tunneling [29] heterostructures has been observed. We also note that due to the small magnitude of the effect, its origin cannot be clearly attributed to the antiferromagnet. Current-induced torque generated by an antiferromagnet has not been experimentally demonstrated so far.

A related issue is whether an antiferromagnet can be influenced by a spin-polarized current injected, for example, from a ferromagnet. Gomonay and Loktev [30] showed that a spin-polarized current can efficiently manipulate the antiferromagnetic order, assuming that the torque generated by the spin-polarized current has on each sublattice the same form as is common in ferromagnets, i.e., $T_j \sim M_j \times (M_j \times p)$ (the so-called antidamping-like torque). Here $j$ is a sublattice index, $M_j$ is the magnetic moment on a sublattice $j$ and $p$ is the direction of the spin-polarization of the current. This form of the torque has been later derived in a more rigorous way. [31, 32] It is instructive to consider why such a torque is effective for manipulating antiferromagnets. The torque can be thought of as generated by an effective magnetic field, such that $T_j \sim M_j \times B_j$ with $B_j = M_j \times p$. In a collinear antiferromagnet such a field is staggered, i.e., alternating in sign between sublattices. It is a general principle that staggered fields can efficiently manipulate antiferromagnets, while uniform fields (such as an external magnetic field) cannot. Unlike the current-induced torque generated by an antiferromagnet, the torque due to the spin-polarized current is expected to be much more robust against disorder. We refer to the article on dynamics in this focused issue for more details on the phenomenology of the time-dependent phenomena induced by the staggered fields.

The torques acting on an antiferromagnet due to injection of a spin-polarized current have been studied in ferromagnet/antiferromagnet bilayers in which the ferromagnet and the antiferromagnet are exchange coupled. The exchange coupling leads to a phenomenon known as exchange bias [33], which is a shift of the ferromagnetic hysteresis loop. It occurs because the antiferromagnet is much less sensitive to magnetic fields than the ferromagnet and thus it pins the ferromagnet in one direction. Several experiments have observed that a current perpendicular to the interface [34,35,36] as well as a current parallel with the interface [37,38,39] influences the exchange bias. This provides indirect evidence that a spin-polarized current can influence antiferromagnets. However, a direct measurement of this effect has not yet been provided. Exchange bias is a complex phenomenon, which depends sensitively on the properties of the interface and still lacks a coherent microscopic description. These experiments therefore cannot pinpoint the exact origin of the observed effect. It is not clear in particular whether the current influences only magnetic moments near the interface (which may be uncompensated) or whether it also influences the magnetic moments in the bulk of the antiferromagnet.

Most of the works discussed so far focused on collinear antiferromagnets. Here the electrical current is not spin-polarized, which limits the spintronics effects that can be present. Recently Železný et al.[40] predicted that in non-collinear antiferromagnets, such as $Mn_3Ir$ and $Mn_3Sn$, the electrical current is spin-polarized. Such antiferromagnets could then generate a spin-transfer torque on another antiferromagnet or a ferromagnet in the same way as a ferromagnet does. Furthermore, robust magnetoresistance effects in metallic or tunneling junctions are also expected to occur, though more research is necessary to show whether the magnitude of these effects could be comparable to ferromagnets. Finally we note that in non-collinear non-coplanar antiferromagnets a Hall-like transport signal can occur. The phenomenon is called a topological Hall effect [41,42] because of its connection to the topology of magnetic textures such as skyrmions. A related non-relativistic phenomenon called a topological SHE has also been identified [43,44,45] in non-collinear antiferromagnets. We refer to the article on topological effects in this focused issue for a more detailed discussion of these Hall phenomena.

The effects we have discussed so far are non-relativistic in origin. In contrast, many of the effects that we will discuss in the following sections are caused by the spin-orbit coupling. This is a relativistic term in the Hamiltonian, which couples the spin and the orbital degree of freedom of an electron. Its significance lies in particular in the fact that it couples spin to the lattice and in this way it lowers the symmetry of the system and can generate a variety of non-equilibrium spin phenomena.

## Anisotropic magnetoresistance

The first and still broadly employed method for electrically detecting a reorientation of the magnetization in a ferromagnet is by using the anisotropic magnetoresistance (AMR):[46] the dependence of the resistance on the direction of the magnetization with respect to current or crystal axes.

The AMR tends to be smaller than the giant or tunneling magnetoresistance, however, it is simpler to detect experimentally since it is a bulk effect and thus does not require complex multilayers. Furthermore, since it is even in magnetization, it is equally present in antiferromagnetic materials.[47] However, until recently, the effect had remained elusive because of the difficulty in controlling the magnetic moment direction in antiferromagnets. Nevertheless, the AMR has now been demonstrated in several antiferromagnets. Marti et al.[48] used antiferromagnetic FeRh for the demonstration of the AMR. This material becomes ferromagnetic when heated and responds to applied magnetic fields. By then cooling back into the antiferromagnetic phase with the field still applied, the antiferromagnetic spin direction can be controlled. Other experiments used antiferromagnets exchange coupled to a ferromagnet, [49, 50] large magnetic fields [51,52] or electrical current [2] (as we discuss in depth later) to manipulate the antiferromagnetic moments. The full functional form of AMR, shown in Fig. 2a, was demonstrated by Kriegner et al. [52] in antiferromagnetic MnTe.

AMR has both longitudinal and symmetric transverse components. Historically, the transverse AMR is sometimes called the planar Hall effect but we avoid this terminology since the true Hall effects correspond to the antisymmetric off-diagonal components of the conductivity tensor. Among those the anomalous Hall effect can also be used for detecting the magnetization reversal in ferromagnets. Unlike AMR, it is odd in magnetization and thus not present in collinear antiferromagnets. It has been, however, demonstrated in non-collinear antiferromagnets [53, 54, 55, 56] (see Fig. 1c). We refer to the article on topological phenomena in this focused issue for more in-depth discussion of this phenomenon.

The AMR is useful for experimental detection of switching of antiferromagnets, however, its small magnitude limits the possible miniaturization and the readout speed in devices. [46] Significantly larger effects can be achieved by using a current tunneling from an antiferromagnet to a nonmagnetic metal, an effect called the tunneling anisotropic magnetoresistance (TAMR) (see Fig. 1d). In antiferromagnets the TAMR was predicted by Schick et al. [47] and subsequently demonstrated in experiments by Park et al. [57] who found a very large TAMR effect in a NiFe/IrMn/MgO/Pt tunneling junction. The effect can exceed 160% at low temperature. In this work, the NiFe ferromagnetic layer sensitive to weak magnetic fields and exchange coupled with the IrMn antiferromagnet was used to rotate the antiferromagnetic moments to produce different resistance states. The NiFe/IrMn exchange-spring effect was, however, not robust enough to persist to room temperature. Ralph et al. [58] reproduced the large low-temperature TAMR and highlighted a strong sample dependence of the effect whose detailed microscopic description is still missing.

An alternative TAMR structure of [Pt/Co]/IrMn/AlO$_x$/Pt with a stronger ferromagnet/antiferromagnet exchange coupling used by Wang et al. [59] allowed for the detection of the effect at room temperature. A weak TAMR signal, not exceeding one percent even at low temperatures, was attributed in this structure to the amorphous AlO$_x$ tunnel barrier. Petti et al. [60] demonstrated that the TAMR effect can also exist in structures which contain no ferromagnet. To switch between two different antiferromagnetic states, the samples were cooled from above the Néel temperature in high magnetic fields. By cooling in perpendicular magnetic fields two states with different resistance were obtained. More research is needed to understand precisely the TAMR mechanisms and to optimize the structures, however, these experiments demonstrate that a large magnetoresistance can in principle exist in antiferromagnetic structures.

Antiferromagnetic insulators have also been used as barriers in tunneling junctions. Wang et al. [61] demonstrated TAMR with an antiferromagnetic insulator. Shvets et al. [62] used an antiferromagnetic insulator as a barrier separating two ferromagnetic layers and showed that apart from the ordinary tunneling magnetoresistance, an additional magnetoresistance due to the antiferromagnet occurs.

## Spin-orbit torque switching

Spin-orbit coupling allows for the generation of a current-induced torque in a magnet without spin-injection from an external polarizer. The spin-orbit torque occurs because in crystals with broken inversion symmetry electrical current generates a non-equilibrium spin-polarization (see Fig. 2b). This effect is known as the Edelstein effect (also called the inverse spin-galvanic effect). [63, 64, 65, 66] In a magnetic material, the current-induced spin-polarization exchange-couples to the equilibrium magnetic moments and thus generates a torque. The Edelstein spin-orbit torque was theoretically proposed [67, 68, 69] and experimentally detected [70, 71] in the ferromagnetic semiconductor GaMnAs and recently also in the room-temperature ferromagnetic metal NiMnSb. [72] For the presence of a net current-induced polarization, the inversion symmetry has to be broken, thus the Edelestein spin-orbit torque can only be used for manipulation of ferromagnets with broken inversion symmetry.

Železný et al. [73, 74] predicted that the Edelstein spin-orbit torque will also occur in antiferromagnets with appropriate symmetry and that it can efficiently manipulate the antiferromagnetic order. The reason for the efficient manipulation is that with appropriate symmetry the current-induced spin-polarization contains a component which is staggered. This staggered component in turn generates a staggered effective magnetic field, which as we have discussed previously, can manipulate the antiferromagnetic order efficiently. Železný et al. [73] calculated the spin-orbit torque for antiferromagnetic $Mn_2Au$, which has the crystal structure shown in Fig. 2c. The nonmagnetic crystal of $Mn_2Au$ has inversion symmetry and, therefore, there is no net current-induced spin-polarizations. The Mn sublattices, however, each have locally broken inversion symmetry. As a consequence there can be a current-induced spin-polarization on each Mn sublattice, but (ignoring the magnetic moments) they have to be precisely opposite. When the magnetic moments are present, the inversion symmetry is broken also globally, however, for the spin-orbit torque in $Mn_2Au$ the local inversion symmetry breaking is more important since the global inversion symmetry breaking only leads to uniform spin-polarization.

The Edelstein spin-orbit torque has advantages compared to the other current-induced torques we have discussed earlier. Since the torque is generated locally, it is not particularly sensitive to disorder [75] and since it is a bulk effect it does not require special heterostructures and thin-film antiferromagnets. On the other hand it only works in antiferromagnets with the appropriate symmetry. For example, for the presence of the field-like Edelstestein spin-orbit torque the two spin-sublattices must occupy inversion-partner lattice sites. The switching of an antiferromagnet via this torque was initially demonstrated by Wadley et al. [2] using epilayers of antiferromagnetic CuMnAs which has a symmetry analogous to $Mn_2Au$. Fig. 2d shows a more recent measurement of this switching phenomenon on CuMnAs grown on Si. In CuMnAs (as well as in $Mn_2Au$) the efficient torque has a field-like character, i.e., the torque is generated by an effective field which is approximately independent of the direction of magnetic moments. Because of the symmetry of the crystal, the field is always perpendicular to the electrical current. [74] Thus, by applying perpendicular current pulses, the magnetic moments in CuMnAs can be switched between two perpendicular directions. In the experiment by Wadley et al. [2] the switching was monitored by the AMR. It was shown that the longitudinal and transverse resistances depend on the direction, amplitude and duration of applied current pulses, in a way which is consistent with the expected Edelstein spin-orbit torque. Further experiments used x-ray photoemission electron microscopy, combined with x-ray magnetic linear dichroism, to directly image the antiferromagnetic domain configuration following the current pulses. [2, 76] These studies confirmed the magnetic origin of the electrical signals due to alignment of antiferromagnetic moments orthogonal to the applied current direction. Recently the spin-orbit torque switching combined with the AMR detection was confirmed experimentally also in sputtered films of $Mn_2Au$. [77]

## Memory devices

Antiferromagnets possess a number of properties that make them highly favourable for memory applications. Like their ferromagnetic counterparts, their magnetic state is inherently non-volatile but with the addition that they are robust to external magnetic fields. The antiferromagnetic spin-sublattices with a compensated magnetic moment give them some intriguing additional benefits. The absence of internal dipolar fields favors multistable antiferromagnetic domain configurations, which can be exploited for integrating memory and logic functionality. The antiferromagnetic exchange is also the origin of the ultrafast reorientation dynamics (in the terahertz regime), as well as making antiferromagnets magnetically "invisible" and enabling denser packing of memory elements.

There have been several demonstrations of antiferromagnetic memory devices based on the concepts we have outlined. The metamagnetic FeRh devices described by Marti et al.[48] demonstrate a memory functionality, where the information bits were represented by perpendicular directions of the antiferromagnetically coupled magnetic moments. The memory states were switched by applying magnetic fields at elevated temperatures, above the ferromagnetic transition, and then cooling into the antiferromagnetic regime. Moriyama et al.[78] used the same approach and material system to demonstrate sequential read and write operations of the antiferromagnetic memory stable over a large number of cycles. In addition to this, the transition to the high-temperature ferromagnetic phase was achieved using the Joule heating of the current line.

The CuMnAs devices described by Wadley et al.[2] demonstrate an antiferromagnetic memory which can be both written (using the spin-orbit torque) and read (using AMR) electrically under ambient conditions (Fig. 3). Olejnik et al.[79] explored the multilevel memory behavior of these devices further and showed that they could be used to write up to thousands of states (Fig. 2e). By applying successive current pulses, progressively higher reproducible resistance states can be reached depending on the number and duration of the pulses. Both pulse-counting and pulse-time-integration functionalities were demonstrated, for pulse lengths ranging from milliseconds down to 250ps, close to the limit of contact current injection with commercial pulse generators (Fig. 2e). The micromagnetic domain origins of these multistable states were imaged by Grzybowski et al.[76] Olejnik et al.[79] also demonstrated their compatibility with conventional microelectronic printed circuit boards (Fig. 3c), and low-temperature-growth and fabrication compatibility of CuMnAs with Si or III-V semiconductors (Fig. 3a).

Recently, switching by ps-pulses of an electrical radiation, combined with the AMR readout, has been demonstrated in the same CuMnAs bit-cells.[80] This shows that a non-contact switching of an antiferromagnet is also feasible and opens a path to the development of antiferromagnetic memories for the THz band. In contrast, feromagnets are limited by the GHz writing speed threshold. For writing speeds up to the GHz scale, weak fields are sufficient for switching ferromagnetic moments and the associated energy cost decreases with increasing speed. Above the GHz threshold, however, the trend reverses, with both the field and the energy cost increasing proportionally to the writing speed. As a result, current-induced spin-orbit torque switching has not been pushed in ferromagnetic memory devices to writing speeds exceeding 5 GHz.[81]

In this review, we have mostly focused on transport phenomena. An exciting possibility is also the control of magnetic order by electric field in magnetic insulators. Such functionality has been demonstrated in multiferroic materials which combine ferroelectricity and antiferromagnetism. See the review article by Sando et al. [82] for description of recent progress on the most commonly used multiferroic material $BiFeO_3$.

## Antiferromagnet/ferromagnet spin Hall structures

When current flows through a material, a spin current appears which is transverse to the charge current. This effect is known as the SHE. [4,3] It originates due to the spin-orbit coupling, which causes the electrons with opposite spin to deflect in opposite directions thus creating a spin current. The inverse effect also exists: when a spin current is injected into a material with spin-orbit coupling a transverse voltage appears. The direct and inverse SHE are of key importance for spintronics since they allow for transforming between charge currents and spin currents.

The SHE can be used to generate a spin-orbit torque and by this to switch a ferromagnet. [83, 84] When a SHE material is interfaced with a ferromagnet and electrical field is applied parallel to the interface a spin current flows into the ferromagnet and generates a torque on the magnetization. It allows, in principle, for a faster and more efficient switching of ferromagnetic layers than spin-transfer torque. Another advantage of the spin-orbit torque in magnetic tunnel junctions is that the large writing current does not flow through the most delicate part of the device which is the tunnel barrier. Note that the SHE mechanism coexists with the Edelstein spin-orbit torque since interfaces break the inversion symmetry of the structure. [4]

The spin-orbit torque due to SHE could also be used to manipulate antiferromagnetic moments [73] (see Fig. 1g). As we have already discussed, spin current injected into an antiferromagnet will generate a torque which can efficiently manipulate the antiferromagnetic order. Such spin current could be injected from a ferromagnet, but the spin current due to SHE could also be used. A spin-orbit torque in a heavy-metal/antiferromagnet configuration has indeed been observed experimentally. [85] However, the difficulty of manipulating the antiferromagnetic order by magnetic fields did not allow for a detailed characterization of the torque that is routinely done in the heavy-metal/ferromagnet bilayers. The switching due to the SHE torque has not been observed so far in antiferromagnets.

In the past, the research of the SHE mostly focused on non-magnetic materials. It is, however, allowed by symmetry in any material, including antiferromagnetic (and other magnetic) materials. The SHE has indeed been found theoretically and experimentally in several antiferromagnets. [86, 87, 88] Instead of the direct SHE, these experiments demonstrated the inverse SHE, which can be detected electrically by injecting a pure spin current into the antiferromagnet. The spin current is generated by a ferromagnet either using precessing magnetization (the so-called spin pumping) as illustrated in Fig. 4b, or alternatively using a heat gradient. An example of DC voltages measured in a spin pumping experiment for antiferromagnets MnX, X=Fe, Pd, Ir and Pt (see Fig. 4a) is given in Fig. 4c. Several of the antiferromagnets showed a large SHE comparable to the commonly used non-magnetic heavy metals. In addition the spin Hall conductivities are highly anisotropic in these antiferromagnetic alloys as has been demonstrated with measurements of epitaxial films with different growth orientation [89]. This behavior is consistent with intrinsic spin Hall effects as determined by first principle calculations. Furthermore, the spin Hall conductivity seems to be susceptible to manipulation via different magnetic field and temperature cycles resulting in different arrangements of the antiferromagnetic spin structure [90].

Since antiferromagnets can have a large SHE it is expected that they can also be used to generate a spin-orbit torque on a ferromagnet. This has been indeed confirmed by several experiments for various antiferromagnets. [85, 91, 89, 92, 90, 93, 94, 95, 96, 97] The observed spin-orbit torque can be very large, comparable to the largest values for non-magnetic heavy metals. Switching of the ferromagnetic layer by the spin-orbit torque has also been demonstrated. [92, 93, 94, 97] Using an antiferromagnet instead of a non-magnetic metal has some unique advantages. For the spin-orbit torque switching it is preferable to use a perpendicularly magnetized ferromagnet since this allows for faster switching and better scalability. However, in such a case a constant in-plane magnetic field has to be applied to achieve deterministic switching. [98] An antiferromagnetic material can be exchange coupled to the ferromagnet (see Fig. 4d) and thus it can provide an internal effective magnetic field acting on the ferromagnet. This allows for switching of the ferromagnet without an external magnetic field. [92, 93, 94, 97] Another remarkable feature that was observed in antiferromagnet/ferromagnet bilayers [92] is a memristor behavior (see Figs. 4e,f), reminiscent of the multi-level spin-orbit torque switching in bulk antiferromagnets (CuMnAs, $Mn_2Au$) discussed above. [2, 77, 79] This behavior is of great importance for neuromorphic computing where it can simulate synapses. It is attributed in the antiferromagnet/ferromagnet bilayers to a progressive current-induced switching of more magnetic domains in the ferromagnet, which are then kept fixed by the exchange coupling with the multi-domain antiferromagnet (see Fig. 4d). A proof-of-concept artificial neural network based on the spin-orbit torque in a PtMn/CoNi structure has been already demonstrated. [99]

## Outlook

The basic memory functionality, including electrical writing and readout has been demonstrated in antiferromagnets. Within a year from the initial demonstration of the spin-orbit torque switching, the electrical pulse length has been reduced from milliseconds to picoseconds, opening a prospect of ultra-fast magnetic memories. The Heusler-compound related CuMnAs antiferromagnet used in the experiments can be epitaxially grown and devices can be microfabricated using III-V or Si substrates, which allows for future integration in semiconductor circuits. The spin-orbit torque switching has also been demonstrated in $Mn_2Au$, which broadens the range of suitable materials to sputtered transition-metal films. A multilevel (memristor) switching is commonly observed in memories made of single-layer antiferromagnets or antiferromagnet/ferromagnet bilayers.

The momentum in the research of antiferromagnetic memories is immense, yet much remains to be done. To fully utilize the potential of antiferromagnets, several key aspects must be better understood: The factors which contribute to the complex domain structures, and their relative importance, need to be elucidated in order to design and engineer devices and material properties with the requisite domain patterns and stability for specific memory applications. The ultra-fast switching needs to be studied in detail. Promising steps have been made towards time-resolved optical detection [100], as discussed elsewhere in this focused issue in the article on opto-spintronics.

The electrical readout signals observed in antiferromagnets at room temperature have so far been relatively small. This could be sufficient for some applications, but larger readout signals would be desirable for high-density fast read-out memory application. Device and material optimization is necessary to extend the large readout signals observed at small temperatures to room temperature. The only electrical switching method that has been demonstrated so far is the bulk Edelstein spin-orbit torque. Switching using the spin-orbit torque due to the SHE could be more efficient and is expected to work in any antiferromagnet, whereas the Edelstein spin-orbit torque requires antiferromagnets with specific symmetry. More material research is necessary to identify other antiferromagnets that could be switched via the Edelstein spin-orbit torque. The spin-orbit torque generated by antiferromagnets on ferromagnets in the antiferromagnet/ferromagnet bilayers is promising, although its origin is not entirely understood. It is commonly attributed to the SHE, but this has not been proven and other effects likely contribute. Of particular interest is understanding how the spin-orbit torque and the SHE depend on the antiferromagnetic order.

# Figures

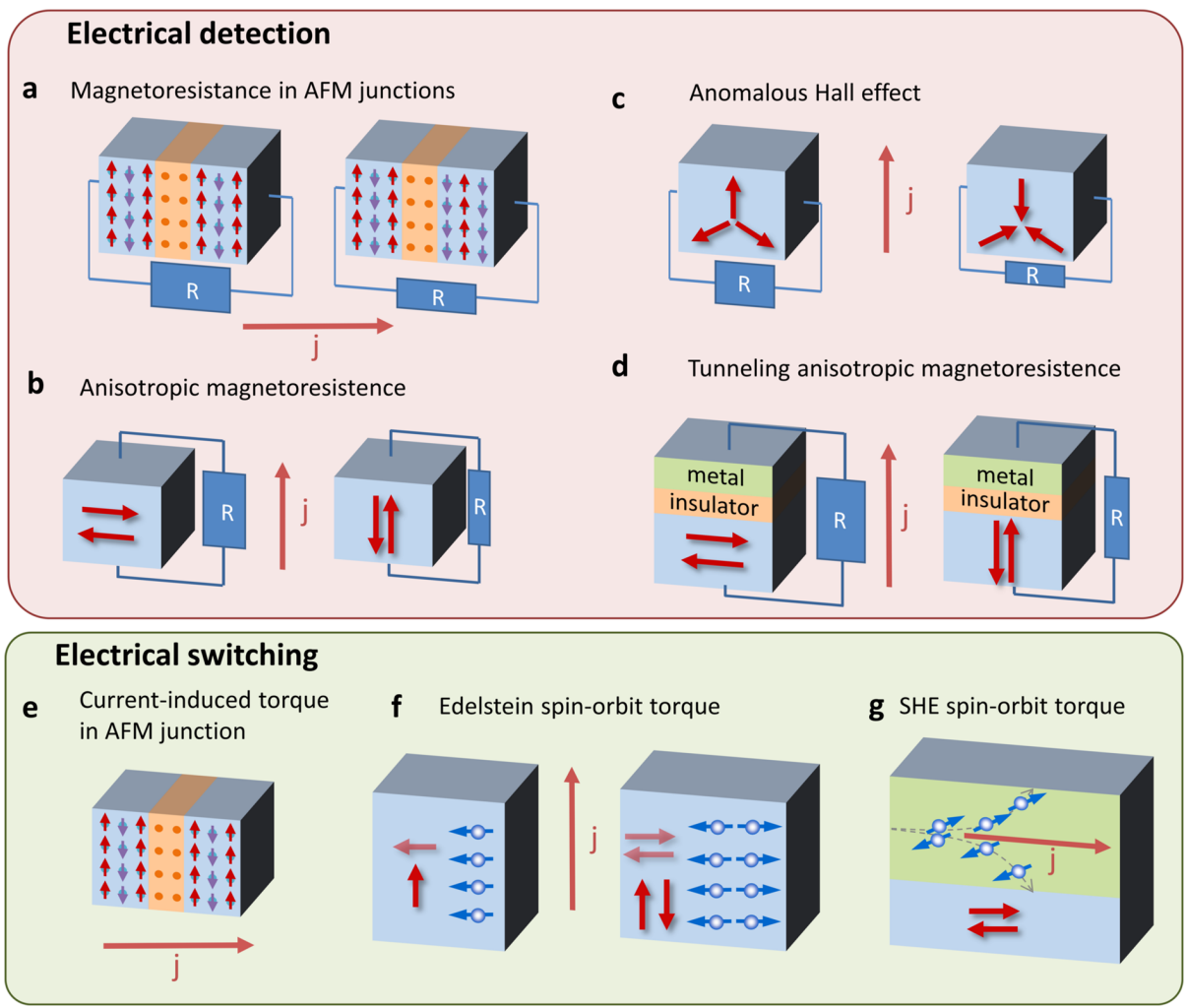

**Figure 1. Illustration of the various concepts proposed for the electrical detection and manipulation of antiferromagnetic order. a**, Magnetoresistance in antiferromagnetic spin valve or tunneling junction. Theoretically proposed, but so far has not been clearly detected experimentally. **b**, Anisotropic

magnetoresistance. Has been experimentally demonstrated in several antiferromagnets. **c**, Anomalous Hall effect. Has been experimentally demonstrated in non-collinear antiferromagnets. **d**, Tunneling anisotropic magnetoresistance. A large readout signal has been experimentally demonstrated. **e**, Current-induced torque in an antiferromagnetic spin valve or tunneling junction. Theoretically proposed, but so far not experimentally demonstrated. **f**, Edelstein spin-orbit torque (left) in a ferromagnet and (right) antiferromagnet. Switching using this torque has been demonstrated experimentally. **g**, SHE spin-orbit torque. The torque in such a device has been experimentally demonstrated, but switching has not been observed yet.

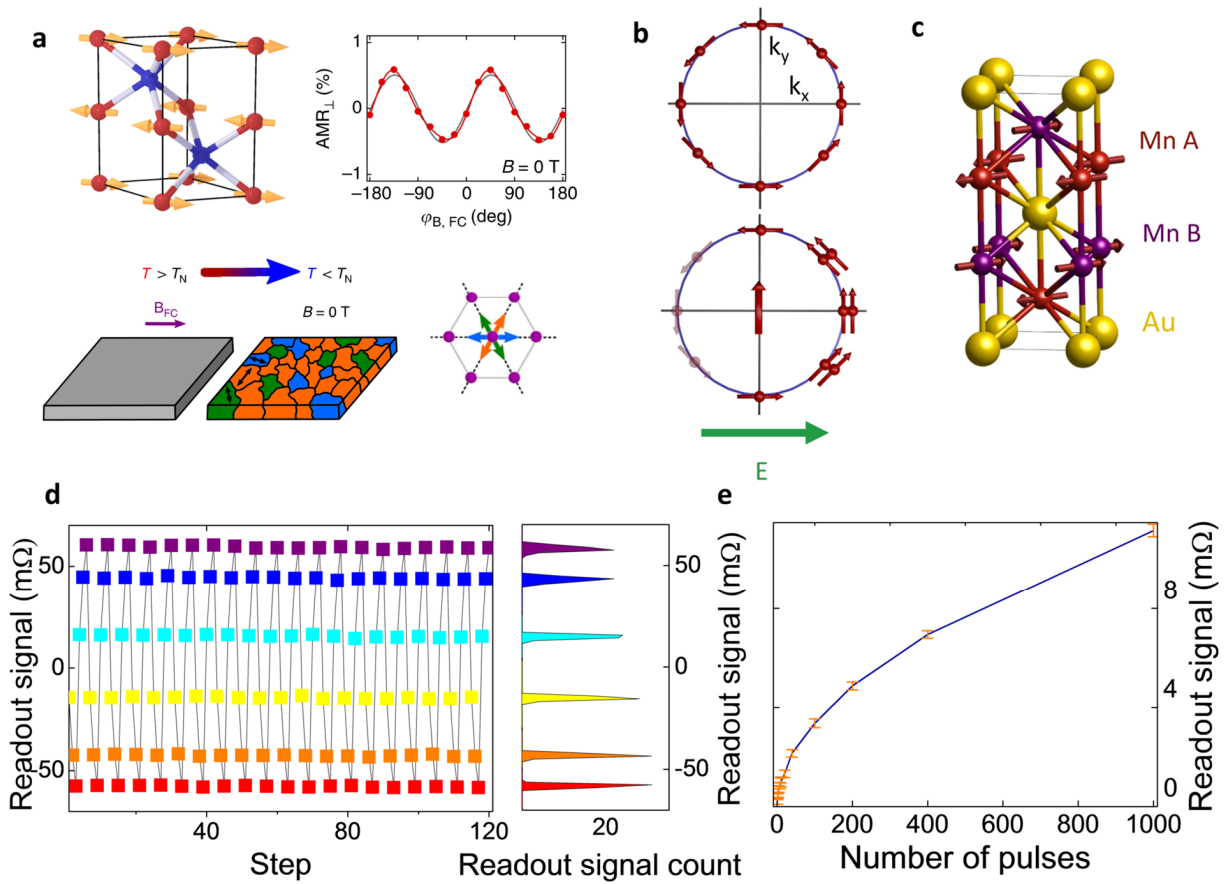

**Figure 2 | AMR and Edelstein spin-orbit torque in antiferromagnets**. **a**, Measurement of AMR in MnTe. Top left: the crystal and magnetic structure of MnTe. Top right: The AMR signal obtained by field-cooling in magnetic fields with various directions. The measurement is done in zero magnetic field. Bottom: Illustration of the field-cooling procedure, which results in a distribution of magnetic domains between the 3 magnetic easy axis (bottom right). From ref. 52. **b**, Illustration of the Edelstein effect. Top: Fermi surface of a system with spin-orbit coupling in equilibrium. Bottom: Applied electric field causes a redistribution of electrons which results in a non-equilibrium spin-polarization. Adapted from Ref. 4. **c**, Crystal and magnetic structure of $Mn_2Au$. **d**, Left: Electrical switching by the Edelstein spin-orbit torque in CuMnAs demonstrating the multi-level stability and reproducibility of the results. For the measurement a series of three pulses was alternatively applied along perpendicular directions. Right:, Histogram of the six different states, obtained from 50 repetitions of the pulse sequence. All

measurements were performed at room temperature. From ref. 79. **e**, Electrical switching by the Edelstein spin-orbit torque in CuMnAs using a series of 250 ps pulses as a function of the number of pulses. The error bars show a standard deviation of 15 repetitions. From ref. 79.

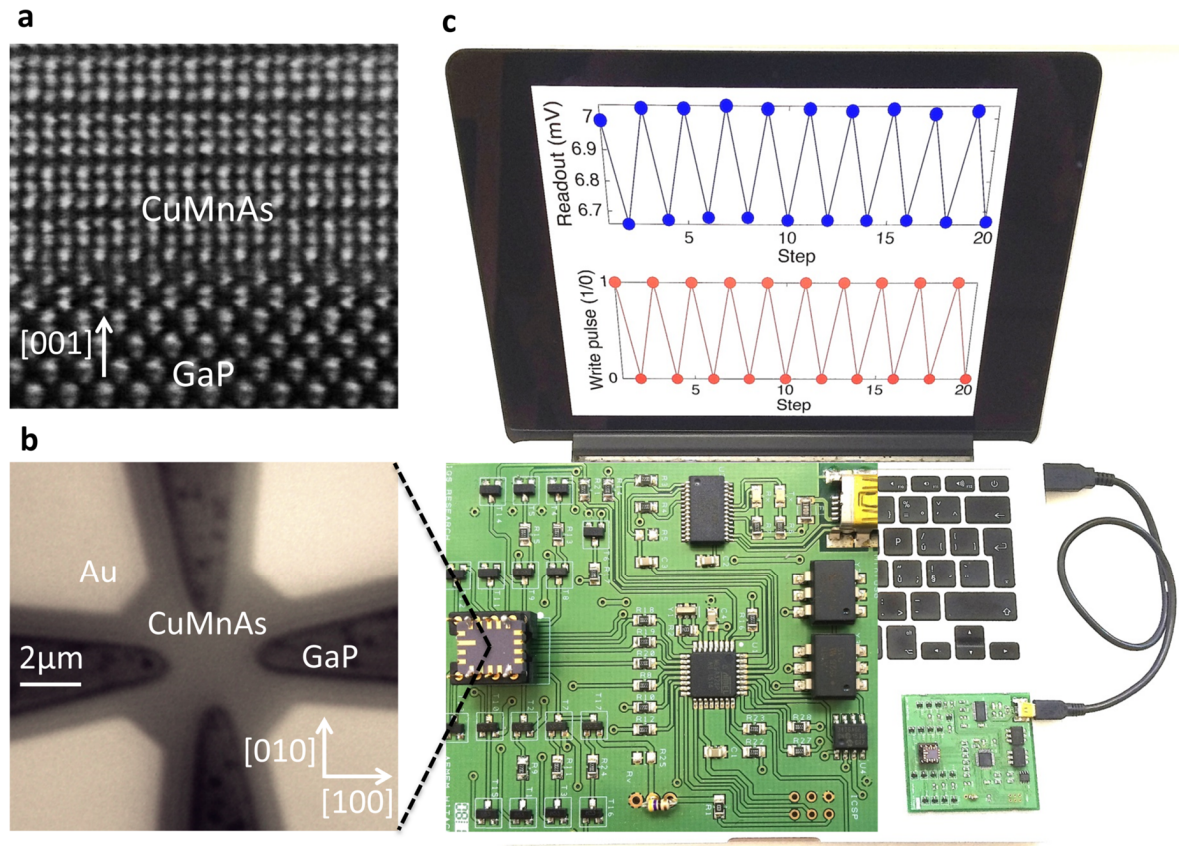

**Figure 3 | Demonstration of microelectronic compatibility of the CuMnAs memory cell**. **a**, Transmission electron miscroscope picture of the epitaxial growth of CuMnAs on GaP. From ref. 79. **b**, The CuMnAs device used for electrical switching. From ref. 79. **c**, The electrical switching of CuMnAs in a USB device. From ref. 79.

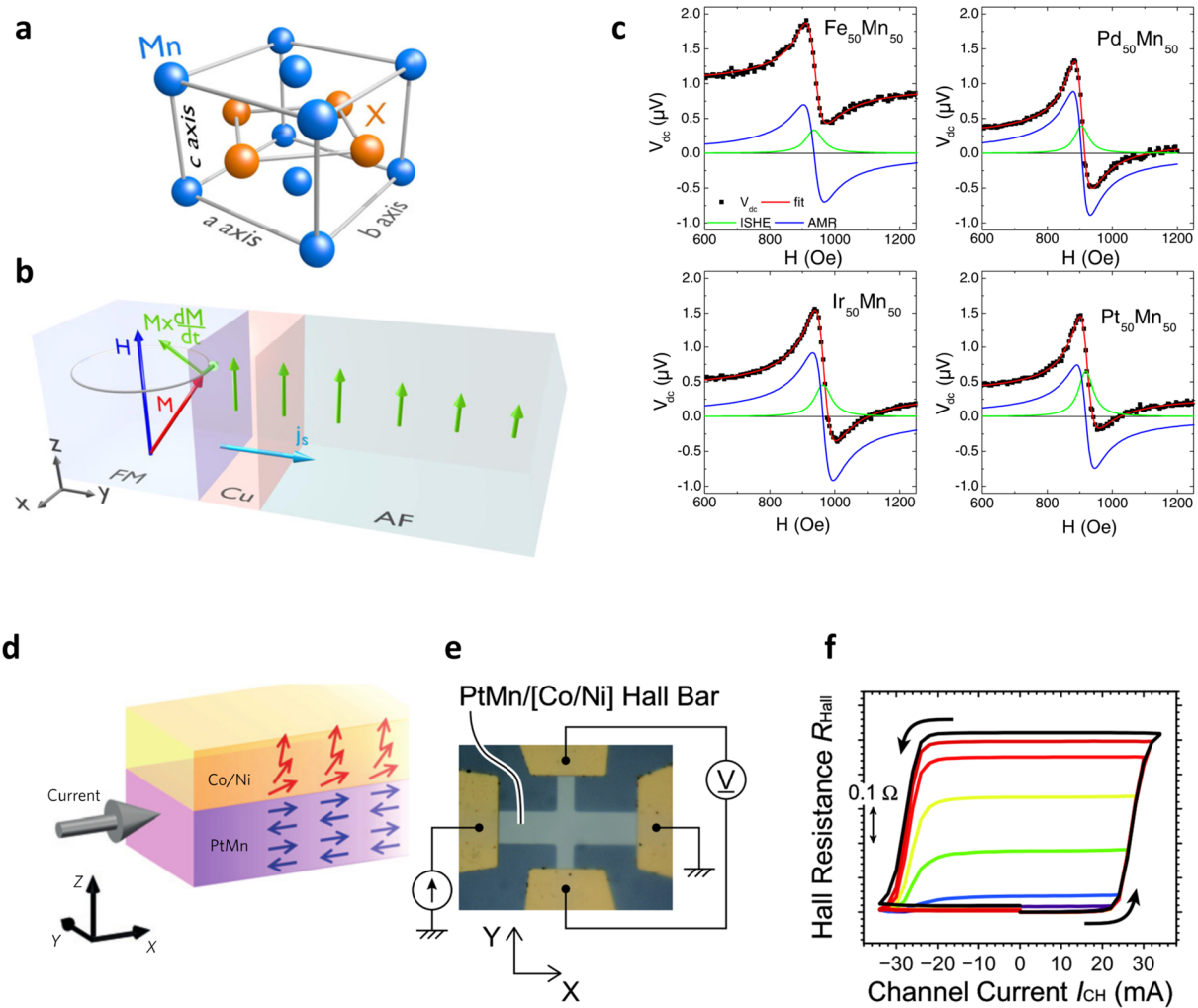

**Figure 4 | SHE and SHE spin-orbit torque in antiferromagnets. a**, Crystal structure of antiferromagnets MnX. From ref. 85. **b**, Schematic of the spin-pumping experiments. Combination of a DC and AC magnetic field induces a precession of the magnetization of the ferromagnetic layer, which causes a spin current flowing in the antiferromagnet. The spin current is then converted to voltage due to the inverse SHE. From ref. 85. **c**, Voltages detected in the spin-pumping experiment as a function of a DC magnetic field for several MnX antiferromagnets. The total signal contains a contribution from inverse SHE and AMR. From ref. 85. **d**, Illustration of the exchange bias in antiferromagnet/ferromagnet bilayers, which leads to a memristor like behaviour of the spin-orbit torque switching From ref. 91. **e**. The experimental setup used for the demonstration of the memristor behaviour. From ref. 98. **f**, Demonstration of the memristor-like behaviour in the PtMn/[Co/Ni] devices. 0.5 s long current pulses with magnitude varying between $-I_{max}$ and $I_{max}$ were applied and the Hall resistance was measured after each pulse. Different values of $I_{max}$ are denoted by different colours. From ref. 98.